\documentclass[reprint,aps,pra]{revtex4-1}

\usepackage{graphicx}
\usepackage{dcolumn}
\usepackage{bm}
\usepackage{amsmath}
\usepackage{amsthm}
\usepackage{amssymb}
\usepackage{float}
\usepackage{hyperref}
\usepackage{color}
\usepackage{epstopdf}
\usepackage{cleveref}
\usepackage[svgnames]{xcolor}
\hypersetup{hidelinks,colorlinks=true,allcolors=DarkBlue}

\newcommand{\ket}[1]{|#1\rangle}
\newcommand{\bra}[1]{\langle#1|}

\newcommand{\Rx}[2]{R_X^{(#1)}\left(#2\right)}
\newcommand{\Ry}[2]{R_Y^{(#1)}\left(#2\right)}
\newcommand{\pot}{\frac{\pi}{2}}
\newcommand{\poe}{{\pi}/{8}}

\begin{document}

\preprint{APS/123-QED}

\title{Single qudit realization of the Deutsch algorithm \\ using superconducting many-level quantum circuits}
\author{E.\,O.\,Kiktenko$^{1,2}$}
\author{A.\,K.\,Fedorov$^{3,1,*}$}
\author{A.\,A.\,Strakhov$^{4}$}
\author{V.\,I.\,Man'ko$^{4,5}$}
\affiliation
{
\mbox{$^{1}$Bauman Moscow State Technical University, Moscow 105005, Russia}
\mbox{$^{2}$Geoelectromagnetic Research Center of Schmidt Institute of Physics of the Earth,}
\mbox{Russian Academy of Sciences, Troitsk, Moscow Region 142190, Russia}
\mbox{$^{3}$Russian Quantum Center, Skolkovo, Moscow 143025, Russia}
\mbox{$^{4}$Moscow Institute of Physics and Technology (State University), Moscow Region 141700, Russia} 
\mbox{$^{5}$P.\,N. Lebedev Physical Institute, Russian Academy of Sciences, Moscow 119991, Russia}
}

\date{\today}

\begin{abstract}
Design of a large-scale quantum computer has paramount importance for science and technologies. 
We investigate a scheme for realization of quantum algorithms using {\it noncomposite} quantum systems, {\it i.e.}, systems without subsystems.
In this framework, $n$ artificially allocated ``subsystems'' play a role of qubits in $n$-qubits quantum algorithms. 
With focus on two-qubit quantum algorithms, we demonstrate a realization of the universal set of gates using a $d=5$ single qudit state.
Manipulation with an ancillary level in the systems allows effective implementation of operators from U$(4)$ group via operators from SU$(5)$ group. 
Using a possible experimental realization of such systems through anharmonic superconducting many-level quantum circuits, 
we present a blueprint for a single qudit realization of the Deutsch algorithm,
which generalizes previously studied realization based on the virtual spin representation  [A.R.\,Kessel et al., {\href{http://dx.doi.org/10.1103/PhysRevA.66.062322}{Phys. Rev. A {\bf 66}, 062322 (2002)}}].

\begin{description}
\item[PACS numbers]
03.65.Wj, 03.65.-w, 03.67.-a
\end{description}
\end{abstract}

\maketitle

\section{Introduction} 

Building of a large-scale quantum computer is one of the most challenging domains of quantum information technologies \cite{Lukin,Manin,Feynman,Ladd,Shor,Grover}.
This generation of computational devices demonstrates a potential to outperform their classical counterparts greatly \cite{Manin,Feynman,Ladd,Shor,Grover}.
Examples include  searching an unsorted database \cite{Grover} as well as integer factorization and discrete logarithm problems \cite{Shor} to name a few. 

From a physical point of view, a quantum computer is an open quantum system with a large number of subsystems, which play a role of information units.
These systems can be realized via a variety of physical platforms.
Quantum states of a composite system are described by the density operator in the abstract Hilbert space being a product,
\begin{equation}\label{eq:HilbertComposite}
	\mathcal{H}=\mathcal{H}_{A}\otimes\mathcal{H}_{B}\otimes\dots\otimes\mathcal{H}_{Z},
\end{equation}
of the Hilbert spaces of the physical subsystems. 
A crucial requirement to such systems as platforms for quantum information processing is scalability with respect to number of qubits \cite{DiVincenzo}. 
Success in scalability of the systems results in increasing the number of subsystems making the problem of achieving a suitable degree of control more and more challenging. 

However, a set of required features for quantum technologies is available not only in composite systems but in noncomposite systems as well \cite{Kessel,Kessel2,Zeilinger,MAManko1,Chernega,MAManko2,MAManko3}.
Recent experimental study of photonic qutrit states demonstrates fundamentally non-classical behavior of noncomposite quantum systems \cite{Zeilinger}.
The idea behind this result dates back to the Kochen--Specker theorem \cite{KochenSpecker},
which provides certain constraints on hidden variable theories, that could be used to explain probability distributions of quantum measurement outcomes.

The Hilbert space of noncomposite systems is arranged in the opposite way to (\ref{eq:HilbertComposite}), however it is equivalent to that mathematically: 
it can be represented in form (\ref{eq:HilbertComposite}), {\it i.e.}, as a product of the Hilbert spaces of {\it abstract} subsystems. 
Investigations of information and entropic characteristics of noncomposite quantum systems \cite{MAManko1,Chernega,MAManko2,MAManko3} 
have confirmed possibilities of their applications in quantum technologies.
Furthermore, a potential gain from the use of noncomposite many-level quantum systems has been demonstrated in 
quantum coin-flipping and bit commitment \cite{QBC},
protocols for quantum key distribution \cite{QKD1,QKD2,QKD3,QKD4}, 
quantum information processing \cite{Kessel,Kessel2,Cereceda,Luo,Gedik,Gedik2}
and clock synchronization algorithms \cite{Tavakoli}.

Remarkably, these studies are supported by substantial progress in experiments with many-level states of
photons \cite{Padua}, 
trapped ions \cite{Hensinger}, 
NMR setups \cite{Gedik2}, 
and superconducting quantum circuits \cite{Katz,Katz2,Gustavsson,Katz3,Ustinov}.

\begin{figure}[t]
\begin{centering}
\includegraphics[width=0.21\textwidth]{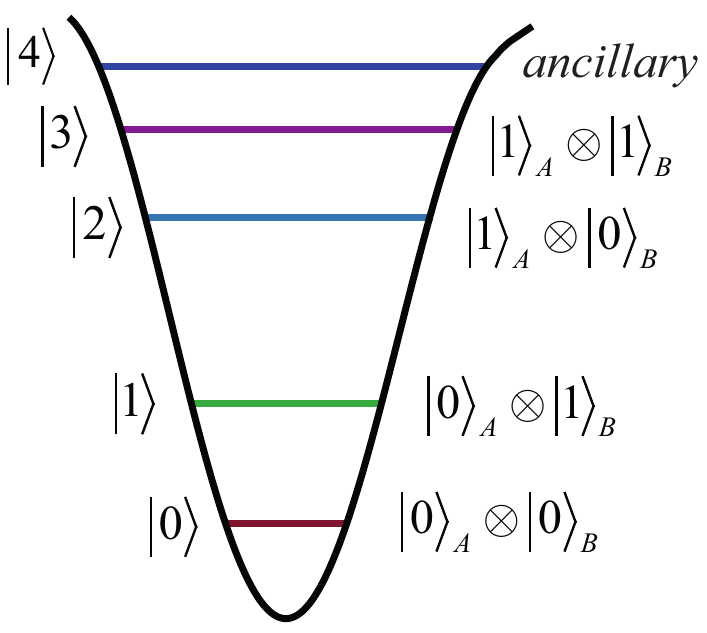}
\end{centering}
\vskip -4mm
\caption
{
Mapping of a five-level quantum system on a two-qubit quantum system.
}
\label{fig:1}
\end{figure}

In the present work, 
we stress on the implementation of quantum algorithms via noncomposite quantum systems with focus on their realization via anharmonic superconducting many-level quantum circuits using addressing to a particular transition.
Our consideration is valid for an arbitrary realized many-level quantum system.
However, we focus on many-level superconducting circuits due to significant progress in their design \cite{Katz,Katz2,Gustavsson,Katz3,Ustinov}.

These advances allow to create highly tunable artificial atomic systems 
with possibilities to reproduce interesting quantum effects \cite{Delsing,Delsing2,Abdumalikov,Wallraff2,Hakonen,Hakonen2,Delsing3,Hakonen3,Lozovik} as well as employ them for quantum computing \cite{You,Wallraff,DiCarlo,Martinis,Wallraff3} and simulation \cite{Wallraff4}.

Recent experiment with a superconducting four-level quantum circuit has explored ``hidden'' two-qubit dynamics \cite{Katz3}.
Therefore, it is interesting to study possibilities of demonstration computational speed-up from single qudit realization of oracle-based algorithms using superconducting many-level circuits.

Here, we consider a qudit state with $d=5$, where four levels are used for storage of two-qubit quantum states and an ancillary fifth level is employed for effective realization of operators from U$(4)$ group via operators from SU$(5)$ group (see Fig. \ref{fig:1}).
We demonstrate that this trick makes it possible to construct the universal set of two-qubit quantum gates consisting of Hadamard, $\pi/8$ and controlled NOT gates \cite{Nielsen}.

The main emphasis of our work is on a single qudit realization of one the first oracle-based quantum algorithm --- the Deutsch algorithm \cite{Deutsch}.
Employment of the ancillary level is a novel feature compared to our previous study \cite{FKMM}, 
where we considered a $d=4$ qudit state and proposed a scheme for Hadamard gates from the universal set only, 
as well as with previously studied realization of the Deutsch algorithm \cite{Kessel2}.
The suggested single qudit realization of the Deutsch algorithm differs from previously studied \cite{Kessel2}, where the operated physical environment allowed to apply arbitrary quantum gates without using of ancillary levels. 

Our paper is organized as follows. 
In Section \ref{universal}, 
we consider a correspondence between a qudit state with $d=5$ and a two-qubit quantum system as well as propose scheme for the universal set of quantum gates for two-qubit algorithms using noncomposite quantum systems.
Using the universal set of quantum gates, we present a realization for a single qudit realization of the Deutsch algorithm in Section \ref{Deutsch}.
We conclude the paper and summarize results in Section \ref{conclusions}.

\section{Universal set of gates}\label{universal}

\begin{table*}[htp] 
\begin{tabular} {| c | c | c | c | } 
\hline 
	Gates 	
	& 
	Action/Control qubit $A$  
	& 
	Action/Control qubit $B$  
	\\ \hline  
	Hadamard gates (\ref{Hadamard}) &
	$
	\frac{1}{\sqrt{2}}
	\begin{bmatrix}
		1 & 0 & 1 & 0 & 0 \\ 
		0 & 1 & 0 & 1 & 0\\
		1 & 0 & -1 & 0 & 0 \\
		0 & 1 & 0 & -1 & 0 \\
		0 & 0 & 0 & 0 &\sqrt{2}
	\end{bmatrix}
	$
	&
	$
	\frac{1}{\sqrt{2}}
	\begin{bmatrix}
		1 & 1 & 0 & 0 & 0 \\ 
		1 & -1 & 0 & 0 & 0 \\
		0 & 0 & 1 & 1 & 0 \\
		0 & 0 & 1 & -1 & 0 \\
		0 & 0 & 0 & 0 & \sqrt{2}
	\end{bmatrix}
	$
	\\  \hline 
	$\pi/8$ gates (\ref{poe})
	&
	$
	\begin{bmatrix}
		1 & 0 & 0 & 0 & 0 \\ 
		0 & 1 & 0 & 0 & 0 \\
		0 & 0 & \exp(i\pi/4) & 0 & 0 \\
		0 & 0 & 0 & \exp(i\pi/4) & 0 \\
		0 & 0 & 0 & 0 & -i
	\end{bmatrix}
	$
	&
	$
	\begin{bmatrix}
		1 & 0 & 0 & 0 & 0 \\ 
		0 & \exp(i\pi/4) & 0 & 0 & 0 \\
		0 & 0 & 1 & 0 & 0 \\
		0 & 0 & 0 & \exp(i\pi/4) & 0 \\
		0 & 0 & 0 & 0 & -i
	\end{bmatrix}
	$
	\\ \hline 
	CNOT gates (\ref{CNOT})
	&
	$
	\begin{bmatrix}
		1 & 0 & 0 & 0 & 0 \\ 
		0 & 1 & 0 & 0 & 0 \\
		0 & 0 & 0 & 1 & 0 \\
		0 & 0 & 1 & 0 & 0 \\
		0 & 0 & 0 & 0 & -1
	\end{bmatrix}
	$
	&
	$
	\begin{bmatrix}
		1 & 0 & 0 & 0 & 0 \\ 
		0 & 0 & 0 & 1 & 0 \\
		0 & 0 & 1 & 0 & 0 \\
		0 & 1 & 0 & 0 & 0 \\
		0 & 0 & 0 & 0 & -1
	\end{bmatrix}
	$
	\\ \hline 
\end{tabular} 
\caption
{
Matrix representation of the universal set of gates: Hadamard, C-NOT, and $\poe$ gates.
The auxiliary level allows effective implementation of operators from U$(4)$ group via operators from SU$(5)$ group, which results in possibility to operate quantum algorithms with accumulation and control for errors. 
} 
\label{tab:representation} 
\end{table*} 

The composite representation of noncomposite quantum $d$-level systems with $d>2$ 
corresponds to any possible mapping of its Hilbert space on a tensor product of several Hilbert spaces, which correspond to abstract subsystems.

In this paper, we consider the five-dimensional Hilbert space of anharmonic superconducting many-level quantum circuit (see Fig. \ref{fig:1}). 
The correspondence between the stationary energy states and two-qubit logic basis can be presented as follows:
\begin{equation}\label{eq:map}
\begin{split}
	&|0\rangle \rightarrow \ket{0}_A\otimes\ket{0}_B, 
	\quad 
	|1\rangle \rightarrow \ket{0}_A\otimes\ket{1}_B, \\
	&|2\rangle \rightarrow \ket{1}_A\otimes\ket{0}_B, 
	\quad 
	|3\rangle \rightarrow \ket{1}_A\otimes\ket{1}_B.\\
	\end{split}
\end{equation}
This mapping resembles the virtual spin representation suggested in Ref. \cite{Kessel}.
We assume that the population of the fifth level is negligible and we keep them in consideration only for the implementation of quantum gates.

Due to above-stated assumptions, the state of the system, written in the original basis, can be presented as
\begin{equation} \label{eq:rho}
	\rho\equiv\rho_{AB}=
	\begin{bmatrix}
		\rho_{00} & \rho_{01} & \rho_{02} & \rho_{03} & 0\\
		\rho_{01}^{*} & \rho_{11} & \rho_{12} & \rho_{13} & 0 \\
		\rho_{02}^{*} & \rho_{12}^{*} & \rho_{22} & \rho_{23} & 0 \\
		\rho_{03}^{*} & \rho_{13}^{*} & \rho_{23}^{*} & \rho_{33} & 0 \\
		0 & 0 & 0 & 0 & 0
	\end{bmatrix},
\end{equation}
while the states of allocated ``subsystems'' $A$ and $B$ turn to have a form
\begin{equation}\label{eq:rhoA}
\begin{aligned}
		\rho_A&=\begin{bmatrix}
		\rho_{00}+\rho_{11} & \rho_{02}+\rho_{13} \\
		\rho_{02}^{*}+\rho_{13}^{*} & \rho_{22}+\rho_{33}
	\end{bmatrix}, \\
		\rho_B&=\begin{bmatrix}
		\rho_{00}+\rho_{22} & \rho_{01}+\rho_{23} \\
		\rho_{01}^{*}+\rho_{23}^{*} & \rho_{11}+\rho_{33}
	\end{bmatrix},
\end{aligned}
\end{equation}
where the matrices are written in their corresponding computational bases.

We assume that our toolbox the system manipulation consists of applying $\theta-$pulses on the transition between arbitrary pair of energy levels. 
In general, it can be done via coupling of a superconducting many-level quantum circuit to an external resonant field \cite{Katz,Katz2,Katz3,Ustinov,Gustavsson}.

The corresponding elementary procedure turns to be rotation around $X$-axis of the ``Bloch sphere'' of the particular two-dimensional Hilbert subspace:
\begin{equation}\label{eq:thetapulse}
	\Rx{jk}{\theta}= \begin{bmatrix}
		\cos({\theta}/{2}) & -i\sin({\theta}/{2}) \\
		-i\sin({\theta}/{2}) & \cos(\theta/{2}) 
	\end{bmatrix}^{(jk)}
	\oplus
	{\mathbb I}_3^{(\overline{jk})},
\end{equation}
where the matrix superscript $j,k\in\{0,1,2,3,4\}$ indicates that it is written in the basis $\{\ket{j},\ket{k}\}$, 
$\oplus$ stands for the direct sum, 
${\mathbb I}_n$ stands for the identity operator in $n$-dimensional Hilbert space and superscript $(\overline{jk})$ indicates that the identity operator acts in the orthogonal complement $\left(\mathrm{Span}\{\ket{j},\ket{k}\}\right)^\bot$, 
then $\Rx{jk}{\theta}$ acts in the whole original five-dimensional Hilbert space.

The appropriate sequence of rotations around $X$-axis results in the effective rotation around $Y$-axis:
\begin{equation}
\begin{split} \label{eq:Ry}
	\Ry{jk}{\theta}&=\begin{bmatrix}
		\cos({\theta}/{2}) & -\sin({\theta}/{2}) \\
		\sin({\theta}/{2}) & \cos(\theta/{2}) 
	\end{bmatrix}^{(jk)}
	\oplus
	{\mathbb I_3}^{(\overline{jk})}\\
	&=\Rx{jl}{\pi}\Rx{kl}{\theta}\Rx{jl}{3\pi},
\end{split}
\end{equation}
where the $l$-th level is one from $(\overline{jk})$.
We note that (\ref{eq:thetapulse}) and (\ref{eq:Ry}) correspond to SU$(5)$ group of unitary operations with the unit determinant.

It is well-known~\cite{Nielsen} that for the case of two-qubit systems the universal set of gates consists of one-qubit Hadamard and $\poe$--gates,
\begin{equation}
	H=\frac{1}{\sqrt{2}}\begin{bmatrix}
		1& 1 \\
		1& -1
	\end{bmatrix}, \quad
	T=\begin{bmatrix} 
		1 & 0 \\
		0 & \exp{(i \pi/4)}
	\end{bmatrix},
\end{equation}
together with two-qubit controlled NOT gates,
\begin{equation}
	U_\mathrm{CNOT}^{(A\rightarrow B)}=\begin{bmatrix}
		1 & 0 & 0 & 0 \\
		0 & 1 & 0 & 0 \\
		0 & 0 & 0 & 1 \\
		0 & 0 & 1 & 0
	\end{bmatrix}, \quad
	U_\mathrm{CNOT}^{(B\rightarrow A)}=\begin{bmatrix}
		1 & 0 & 0 & 0 \\
		0 & 0 & 0 & 1 \\
		0 & 0 & 1 & 0 \\
		0 & 1 & 0 & 0
	\end{bmatrix}.
\end{equation}
Here, arrows in superscripts indicate which qubit is the control one in this operation. 

In our setup, we can implement the Hadamard gates on particular qubits $A$ and $B$ as follows:
\begin{equation}\label{Hadamard}
\begin{split}
	&H^{(A)}{=}\Rx{23}{\pi}\Rx{12}{\frac{7\pi}{2}}\Rx{03}{\frac{7\pi}{2}}\Rx{23}{\pi} \\
	&H^{(B)}{=}\Rx{13}{\pi}\Rx{12}{\frac{7\pi}{2}}\Rx{03}{\frac{7\pi}{2}}\Rx{13}{\pi}.
\end{split}
\end{equation}

In turn, the $\pi/8$ gates, acting on a particular qubit, have the following form:
\begin{equation}\label{poe}
\begin{split}
	T^{(A)}&=\Ry{24}{\frac{7\pi}{2}}\Rx{24}{\frac{7\pi}{2}}\Ry{24}{\pot}\times\\
	&\times\Ry{34}{\frac{7\pi}{2}}\Rx{34}{\frac{7\pi}{2}}\Ry{34}{\pot} \\
	T^{(B)}&=\Ry{14}{\frac{7\pi}{2}}\Rx{14}{\frac{7\pi}{2}}\Ry{14}{\pot}\times\\
	&\times\Ry{34}{\frac{7\pi}{2}}\Rx{34}{\frac{7\pi}{2}}\Ry{34}{\pot}.
\end{split}
\end{equation}
It easy to see that the determinant of four-dimensional operators $T\otimes\mathbb{I}_2$ and $\mathbb{I}_2 \otimes T$ is equal to $i$, 
so they cannot be represented as a sequence of operators from SU$(4)$.
This fact is the crucial reason for the introduction of the ancillary level.
In the case of $\pi/8$ gates, it accumulates the phase $-i$ to obtain a unit determinant and makes it possible to realize the desired operation.

Finally, the CNOT gates can be implemented as
\begin{equation}\label{CNOT}
\begin{split}
	\mathcal{U}_\mathrm{CNOT}^{(A\rightarrow B)}&=\Ry{23}{\pi}\Rx{34}{2\pi} \\
	\mathcal{U}_\mathrm{CNOT}^{(B\rightarrow A)}&=\Ry{13}{\pi}\Rx{34}{2\pi},
\end{split}
\end{equation}
where again we have to address the ancillary level to accumulate additional phase.

The full form of resulting operators is presented in Table \ref{tab:representation}.

\section{Deutsch algorithm}\label{Deutsch}

\begin{figure*}[htbp]
\includegraphics[width=2\columnwidth]{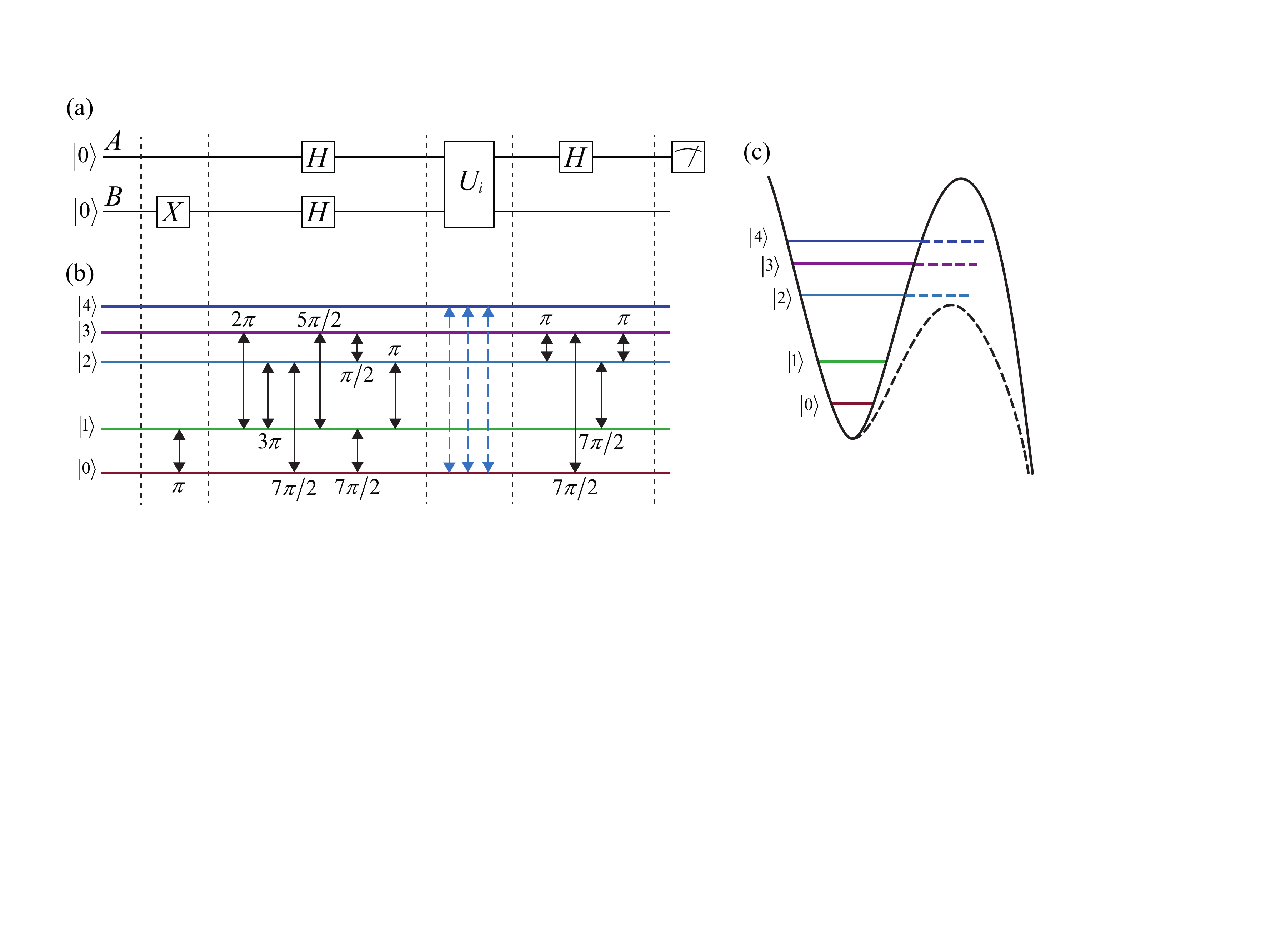}
\caption
{
Blueprint for a single $d=5$ qudit realization of the Deutsch algorithm.
In (a) the Deutsch algorithm's quantum circuit.
In (b) a corresponding sequence of $\theta$-pulses for an anharmonic superconducting many-level quantum circuit used as a platform for qudit.
Blue dashed arrows denote a corresponding sequence from~(\ref{eq:Utilde}).
The answer to the original question about the type of the function $f_j$ could be obtained using a coarse-grained measurement of the energy level: 
one needs to know whether it is higher than the energy of $\ket{1}$ or not.
In (c) a proposal for the readout scheme based on the variation of the potential and checking whether the tunneling effect takes place.
Similar readout scheme for an anharmonic four-level superconducting quantum circuit has been used in Ref. \cite{Katz3}.}
\label{fig:Deutsch}
\end{figure*}

Let us consider a ``black box'', which implements a Boolean function of one argument $f$.
In fact, there are only four possible variants of such function: 
\begin{equation}
	\begin{aligned}
		f_1(0)=0, 
		\qquad 
		f_1(1)=0; \\
		f_2(0)=1,
		\qquad
		f_2(1)=1; \\
		f_3(0)=0,
		\qquad
		f_3(1)=1;	\\
		f_4(0)=1,
		\qquad
		f_4(1)=0.
	\end{aligned}
\end{equation}
We note that $f_1$ and $f_2$ return the same value for all possible inputs and are called {\it unbalanced} functions, 
while $f_3$ and $f_4$ returns value 1 for half of the inputs, and 0 for the other half, and are called {\it balanced} ones.
The considered task is to determine whether the unknown function $f$ is balanced or not by minimal amount of queries to the black box.
Clearly, in a classical domain one need at least two queries to cope with this task.

In a quantum domain, the same problem is formulated using a set of two-qubit quantum gates $\{U_i\}_{i=1}^4$, where each gate performs the following operation
\begin{equation}\label{eq:U}
	U_i|x\rangle \otimes |y\rangle=|x\rangle\otimes |y \mathrm{~XOR~} f_i(x)\rangle, \quad x,y \in \{0,1\}.
\end{equation}
The problem is to determine whether, for the given gate $U_j$ the corresponding function $f_j$ is balanced or not.
The Deutsch algorithm \cite{Deutsch} copes with this problem within only one query.
It turns out that the following sequence of quantum gates gives:
\begin{multline}
	(H \otimes \mathbb{I})U_j(H \otimes H)(\mathbb{I} \otimes X)\ket{0}\otimes\ket{0} \\
	=(-1)^j\ket{f_j(0)\mathrm{~XOR~} f_j(1)}\otimes\ket{-},
\end{multline}
where we introduce the following notations:
\begin{equation}
	X=\ket{0}\bra{1}+\ket{1}\bra{0}, \quad 
	\ket{-}=2^{-1/2}(\ket{0}-\ket{1})
\end{equation}
One can see that the resulting value of the first qubit contains the answer for the task. 
Indeed, it is 1 for balanced function functions and 0 otherwise.

In the framework of the considered the two-qubit mapping~(\ref{eq:map}) the counterparts of gates~(\ref{eq:U}) can be realized as follows:
\begin{equation}\label{eq:Utilde}
	\begin{aligned}
		\mathcal{U}_1&=\mathbb{I}_5, \\
		\mathcal{U}_2&=\Ry{02}{2\pi}\Ry{01}{\pi}\Ry{23}{\pi}, \\
		\mathcal{U}_3&=\mathcal{U}_\mathrm{CNOT}^{(A\rightarrow B)}=\Ry{23}{\pi}\Rx{34}{2\pi}, \\
		\mathcal{U}_4&=\Ry{01}{\pi}\Rx{14}{2\pi}.
	\end{aligned}
\end{equation}

It can be directly checked that the acting of $\mathcal{U}_j$ given by (\ref{eq:Utilde}) on the density matrix in form (\ref{eq:rho}) 
after mapping (\ref{eq:map}) gives the same result as acting of $U_j$ (\ref{eq:U}) on corresponding two-qubit density matrix.

Depending on particular transformation (\ref{eq:Utilde}), one can obtain the following state:
\begin{equation}
	\ket{\psi_j}=H^{(A)}\mathcal{U}_jH^{(AB)}\Ry{01}{\pi}\ket{0}
\end{equation}
with
\begin{equation} \label{eq:psis}
	\begin{split}
		\ket{\psi_1}=\frac{i}{\sqrt{2}}(\ket{0}-\ket{1}), 
		\quad 	
		\ket{\psi_2}=\frac{i}{\sqrt{2}}(\ket{1}-\ket{0}),
		\\	
		\ket{\psi_3}=\frac{i}{\sqrt{2}}(\ket{2}-\ket{3}), 
		\quad 	
		\ket{\psi_4}=\frac{i}{\sqrt{2}}(\ket{3}-\ket{2}),
	\end{split}
\end{equation}
where $H^{(AB)}$ stands for Hadamard gates acting on both subsystems $A$ and $B$. 
This operation can be implemented as follows:
\begin{equation}
\begin{split}
	&H^{(AB)}\equiv H^{(A)}H^{(B)}=
	\Rx{12}{\pi}\Rx{23}{\frac{\pi}{2}}
	\times\\
		&~~~~~\times\Rx{01}{\frac{7\pi}{2}}\Rx{13}{\frac{5\pi}{2}}\Rx{02}{\frac{7\pi}{2}}\\
		&~~~~~\times\Rx{12}{3\pi}\Rx{13}{2\pi}.
\end{split}
\end{equation}
A complete scheme for a single qudit realization of the Deutsch algorithm is presented in Fig. \ref{fig:Deutsch}.

By considering the set of states in (\ref{eq:psis}), one can conclude that the answer to the original question whether the function $f_j$ is balanced or not could be obtained by coarse-grained measurement of energy level.
Indeed, one needs to know whether it is higher than the energy of $\ket{1}$ or not.
It can be performed by variation of the potential and checking whether the tunneling effect takes place (Fig.~\ref{fig:Deutsch}c).
Similar experimental setup for the readout scheme has been used in Ref. \cite{Katz3}.

\section{Conclusion}\label{conclusions}

In the present Letter, we used the correspondence between $d=5$ qudit states and ``two-qubit'' quantum systems with an ancillary level given by (\ref{eq:map}) to present single qudit schemes for the universal set of two-qubit gates:
Hadamard (\ref{Hadamard}), $\pi/8$ gates (\ref{poe}), and CNOT gates (\ref{CNOT}).
In our framework, an ancillary fifth level in the systems allowed us to implement operators from U$(4)$ group via operators from SU$(5)$ group.

We suggested a scheme for a single $d=5$ qudit realization of the Deutsch algorithm using an anharmonic four-level superconducting quantum circuit as a platform and applying $\theta-$pulses on the transition between arbitrary pair of energy levels as basic operation for realizing gates.
In our scheme, a standard way to readout based on the variation of the potential can be implemented. 
It is interesting to study possible realization of another class of quantum algorithms and investigate a potential gain from using of noncomposite quantum systems. 

\section*{Acknowledgments} 

We thank K. Shulga and E. Glushkov for helpful comments.   
We acknowledge support from the Dynasty Foundation, the Council for Grants of the President of the Russian Federation (grant SP-961.2013.5, E.O.K.), and the RFBR.

\vspace{\baselineskip}

*Corresponding author: {\href{mailto:akf@rqc.ru}{akf@rqc.ru}}

\end{document}